\documentclass[conference]{IEEEtran}
\IEEEoverridecommandlockouts
\usepackage{cite}
\usepackage{amsmath,amssymb,amsfonts}
\usepackage{algorithmic}
\usepackage{graphicx}
\usepackage{textcomp}
\usepackage{xcolor}

\usepackage{amsmath,graphicx}
\usepackage{cite}
\usepackage{amssymb,amsfonts,amsthm,bm} 

\usepackage{booktabs}
\usepackage{graphicx}
\usepackage{multirow}
\usepackage{stfloats}
\usepackage{array} 
\usepackage{caption}

\def\BibTeX{{\rm B\kern-.05em{\sc i\kern-.025em b}\kern-.08em
    T\kern-.1667em\lower.7ex\hbox{E}\kern-.125emX}}

\begin{document}

\title{A deep representation learning speech enhancement method using $\beta$-VAE 
\thanks{This  work  was partly supported  by   Innovation  Fund Denmark (Grant No.9065-00046).}
}

\author{\IEEEauthorblockN{Yang Xiang$^{ \star\dagger}$, Jesper Lisby Højvang$^{\dagger}$, Morten Højfeldt Rasmussen$^{ \dagger}$, Mads Græsbøll Christensen$^{\star}$}
\IEEEauthorblockA{$^{\star}$ Audio Analysis Lab, CREATE, Aalborg University, Aalbory, Denmark  \{yaxi,mgc\}@create.aau.dk \\
$^{\dagger}$ Capturi A/S, Aarhus, Denmark  \{jlh,mhr\}@capturi.com}
}
\maketitle

\begin{abstract}
In previous work, we proposed a variational autoencoder-based (VAE) Bayesian permutation training speech enhancement (SE) method (PVAE) which indicated that the SE performance of the traditional deep neural network-based (DNN) method could be improved by deep representation learning (DRL). Based on our previous work, we in this paper propose to use $\beta$-VAE to further improve PVAE's ability of representation learning. More specifically, our $\beta$-VAE can improve PVAE's capacity of disentangling different latent variables from the observed signal without the trade-off problem between disentanglement and signal reconstruction. This trade-off problem widely exists in  previous $\beta$-VAE algorithms. Unlike the previous $\beta$-VAE algorithms, the proposed $\beta$-VAE strategy can also be used to optimize the DNN's structure. This means that the proposed method can not only improve PVAE's SE performance but also reduce the number of PVAE training parameters. The experimental results show that the proposed method can acquire better speech and noise latent representation than PVAE. Meanwhile, it also obtains a higher scale-invariant signal-to-distortion ratio, speech quality, and speech intelligibility.    
\end{abstract}

\vspace{-0.2cm}

\section{Introduction}

The aim of speech enhancement (SE) is to remove background noise from the observed speech signal. In general, SE is mainly used to reduce the word error rate of the automatic speech recognition system \cite{li2014overview} or improve speech quality and intelligibility for human listening \cite{wang2018supervised}. Recently, with the wide application of online meeting systems, SE is required to reduce the WER for accurate live caption when providing high-quality speech audio under various complex noise conditions \cite{eskimez2021human}. Thus, SE research is becoming more and more challenging.

During the past decades, many single-channel SE algorithms have been developed, including signal subspace methods \cite{jensen2015noise},  non-negative matrix factorization methods \cite{xiang2021novel,xiang2020nmf}, and codebook-based methods \cite{kavalekalam2018model}. In recent years, deep neural networks (DNN) have shown great potential for SE \cite{wang2018supervised,xu2014regression,wang2014training, wang2021compensation,luo2019conv, xiang2020parallel,hu20g_interspeech, li2021icassp}  because DNNs can use a non-linear process to model complex high-dimensional signals, which is more reasonable in practical applications \cite{bengio2013representation}. Thus, DNN-based methods usually have a better SE performance than these previous linear models\cite{jensen2015noise,xiang2021novel,xiang2020nmf,kavalekalam2018model}.

However, most of the regression-based SE algorithms  \cite{wang2018supervised,xu2014regression,wang2014training, wang2021compensation}  do not consider applying DNNs to obtain better speech representations when conducting SE. Instead, they usually use DNN to directly predict pre-defined targets for SE \cite{wang2018supervised}. Although this approach can avoid inaccurate assumptions \cite{xu2014regression}, it cannot ensure that these methods always work in environments with complex noise  \cite{wang2018supervised}. In general, deep representation learning (DRL) is important for DNN because DRL can obtain good signal representations in an unsupervised way and can, potentially, improve DNN's ability to extract useful information in complex environments \cite{xie2021disentangled,bengio2013representation}. Additionally, a better signal representation usually leads to better predictions for DNNs \cite{bengio2013representation}. Thus, DRL has a huge potential for DNN-based SE algorithms and makes them more robust. Moreover, the lack of a good DRL strategy may cause poor generalization of DNN-based SE algorithms \cite{wang2018supervised,bengio2013representation}. A good DRL algorithm can also disentangle various latent representations \cite{bengio2013representation} of speech signals (e.g., speaker and phoneme information), which can also help DNN achieve a better SE performance.

Recently, to improve traditional DNN's generalization ability, DRL-based SE algorithms are proposed  \cite{leglaive2018variance,bando2018statistical,leglaive2019semi,carbajal2021guided,fang2021variational, carbajal2021disentanglement}. The basic idea of these methods is that they use a variational autoencoder (VAE) \cite{kingma2013auto} to learn speech representations when modeling speech, and apply a non-negative matrix factorization (NMF) to model noise. VAE is a DRL model and can perform efficient approximate posterior inference. Additionally, VAE can also learn the probability distribution of complex data. Thus, VAE is suitable for various speech generative tasks \cite{kingma2013auto,kim2021conditional,zhang2021visinger}. These VAE-based algorithms can effectively improve DNN's generalization ability, but they have difficulty obtaining good speech representations from the observed signal because they cannot disentangle speech representations from other latent representations \cite{bengio2013representation,leglaive2018variance,bando2018statistical,leglaive2019semi,carbajal2021guided,fang2021variational, carbajal2021disentanglement}. This causes the need to use a linear NMF to model noise, so their noise modeling ability is limited compared with these non-linear DNN-based methods \cite{kingma2013auto}. And their SE performance is not always satisfactory in a complex noisy environment \cite{bando2018statistical}.

To obtain a better speech representation from the observed signal, a novel VAE-based SE method (named PVAE) is proposed \cite{xiang2022bayesian}. This method applies an unsupervised method to learn signal representations and derives a novel VAE lower bound, which ensures that VAE can disentangle different latent variables from the observed signal. Compared to the previous VAE-based SE algorithms, PVAE can use non-linear DNNs to model noise, which improves the noise modeling ability. Additionally, this method can adopt various DNN structures \cite{wang2018supervised}, so the DNN-based SE algorithms \cite{wang2018supervised} can be directly optimized by PVAE. This is not achieved by VAE-NMF-based algorithms \cite{leglaive2018variance,bando2018statistical,leglaive2019semi,carbajal2021guided,fang2021variational, carbajal2021disentanglement}. The experimental results \cite{xiang2022bayesian} indicate that the SE performance of the traditional DNN-based methods can be improved by introducing this PVAE-based DRL algorithm.

Inspired by previous works, in this paper we propose a novel $\beta$-VAE strategy to improve PVAE's representation learning and disentangling performance \cite{bengio2013representation} with fewer DNN parameters. $\beta$-VAE \cite{higgins2016beta, burgess2018understanding} is originally designed to push VAE to learn a more efficient latent representation of the data, which is disentangled if the data contains at least some underlying factors of variation \cite{higgins2016beta}. However, in general, $\beta$-VAE has a trade-off problem \cite{burgess2018understanding}. A better disentanglement within the latent representations usually causes worse signal reconstruction. In this work, based on the VAE's application in SE \cite{xiang2022bayesian}, we propose a strategy to address this trade-off problem to obtain better speech and noise representation. As a result, our $\beta$-VAE can improve disentangling and representation performance without signal reconstruction loss. Moreover, the proposed $\beta$-VAE can also optimize the neural network structure of the original PVAE. This means that the proposed $\beta$-VAE (named $\beta$-PVAE) can possibly achieve a better SE performance with fewer training parameters compared to PVAE.

\section{Related work}

\begin{figure}[!tbp]
  \centering
  \setlength{\abovecaptionskip}{0.1cm}
  \centerline{\includegraphics[scale=0.6]{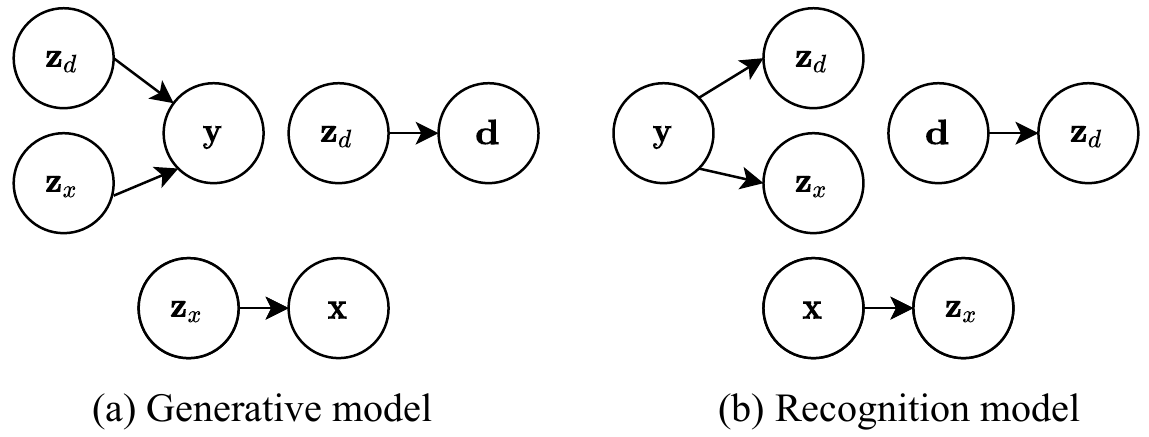}}
  \setlength{\belowcaptionskip}{3pt}
  \caption{Graphical illustration of the proposed signal model.}
  \label{fig:Bayesian_model}
  \vspace{-0.7cm}
\end{figure}

{\bf Signal Model:} in an additive noisy environment, using the short-time Fourier transform, the observed signal $y_{f,n} \in \mathbb{C}$, speech signal $x_{f,n} \in \mathbb{C}$, and noise $d_{f,n} \in \mathbb{C}$ can be written as 
\begin{equation}
\small
  y_{f,n} = x_{f,n}+d_{f,n},
  \label{time_noisy_model}
\end{equation}
where frequency bin $f \in [1, F]$ and time frame index $n \in [1, N]$. $N$ and $F$ denote the number of time frames and frequency bins, respectively. Their log-power spectrum (LPS) vector \cite{xu2014regression} at each frame can be represented as $\mathbf{y}$, $\mathbf{x}$, and $\mathbf{d}$, respectively, where we omit the frequency and time frame index for simplicity. In \cite{xiang2022bayesian}, we assume that $\mathbf{y}$ is generated from a random process involving the speech latent variables $\mathbf{z}_x \in {\mathbb{R}}^L$ and the noise latent variables $\mathbf{z}_d \in {\mathbb{R}}^L$. $L$ is the dimension of latent variables. The latent variables $\mathbf{z}_x$ and $\mathbf{z}_d$ {are} independent. Similarly, $\mathbf{x}$ and $\mathbf{d}$ are independently generated by $\mathbf{z}_x$ and $\mathbf{z}_d$, respectively. Fig.~\ref{fig:Bayesian_model}(a) shows the generative process. In \cite{xiang2022bayesian}, it is assumed that $\mathbf{z}_x$ and $\mathbf{z}_d$ can be estimated from speech and noise posterior distributions $p(\mathbf{z}_x|\mathbf{x})$ and $p(\mathbf{z}_d|\mathbf{d})$, respectively, and that they can also be estimated from the noisy speech posterior distributions $p(\mathbf{z}_x|\mathbf{y} )$ and $p(\mathbf{z}_d|\mathbf{y})$. To disentangle latent variables, we assume that $p({\mathbf{z}_x},{\mathbf{z}_d}|\mathbf{y})=p(\mathbf{z}_x|\mathbf{y})p(\mathbf{z}_d|\mathbf{y})$. Although this assumption is not always accurate in practical environments, it simplifies derivations, and helps us obtain a better signal model. Additionally, its effect towards signal estimation is not significant  \cite{xiang2022bayesian} (related analysis will be also given in Section IV). Fig.~\ref{fig:Bayesian_model}(b) shows the recognition process.

{\bf VAE and $\beta$-VAE:} the original VAE \cite{kingma2013auto} defines a probabilistic generative process between the observed signal and its latent variables, and provides a principled method to jointly learn latent variables, generative and recognition models. The generative and recognition models are jointly trained by maximizing the evidence lower bound \cite{kingma2013auto}
\begin{equation}
\small
 \begin{aligned}
   &  {\mathbb E_{{\bf {y}} \sim p(\bf {y})}[\log q(\bf {y})]} \ge -{\mathcal{L}}_n, \\
   & {\mathcal{L}}_n = {\mathbb E_{{\bf {y}} \sim p(\bf {y})}} \left[D_{KL}\left({p({\bf z}_y|{\bf{y}}))}||{q({{\bf z}_y})}\right)\right] \\
   & \quad \quad - {\mathbb E_{{\bf {y}} \sim p(\bf {y})}} \left[ {\mathbb E_{{{\bf {z}}_y} \sim p({{\bf {z}}_y}|\bf {y})}}\left[\log {q({\bf{y}}|{\bf z}_y)} \right]\right],
   \end{aligned}
  \label{ELBO}
\end{equation}
where $D_{KL}(||)$ denotes the Kullback-Leibler (KL) divergence. $\mathbf{z}_y \in {\mathbb{R}}^L$ is the noisy latent variable. Maximizing this lower bound is equivalent to minimizing ${\mathcal{L}}_n$. 

$\beta$-VAE \cite{higgins2016beta} is a modification of the original VAE framework, which introduces an adjustable hyperparameter $\beta$ in the KL divergence term: 
\begin{equation}
\small
 \begin{aligned}
   & {\mathcal{L}}_n = \beta{\mathbb E_{{\bf {y}} \sim p(\bf {y})}} \left[D_{KL}\left({p({\bf z}_y|{\bf{y}}))}||{q({{\bf z}_y})}\right)\right] \\
   & \quad \quad - {\mathbb E_{{\bf {y}} \sim p(\bf {y})}} \left[ {\mathbb E_{{{\bf {z}}_y} \sim p({{\bf {z}}_y}|\bf {y})}}\left[\log {q({\bf{y}}|{\bf z}_y)} \right]\right].
   \end{aligned}
  \label{beta_ELBO}
\end{equation}
In general, $\beta > 1$ results in more disentangled latent representations \cite{higgins2016beta}. Higher values of $\beta$ can encourage learning a more disentangled representation. However, $\beta$-VAE usually has a trade-off problem between the latent representation disentanglement and signal reconstruction.

{\bf Bayesian permutation training VAE (PVAE) for SE:} Although the VAE-based algorithms \cite{kingma2013auto, higgins2016beta} can learn signal representations and disentangle latent representations in a self-supervised way, their performance is limited when disentangling desired latent representations for SE application. Therefore, a Bayesian permutation training VAE (PVAE) \cite{xiang2022bayesian} is proposed for SE. PVAE is a semi-supervised DRL method, which introduces multiple latent variables in VAE and disentangles them in a semi-supervised way. Fig.~\ref{fig:Bayesian_DNN} shows the PVAE framework. It can be seen that PVAE includes three VAE structures: clean speech VAE (C-VAE), noise VAE (N-VAE), and noisy VAE (NS-VAE). C-VAE and N-VAE are trained without supervision to obtain speech and noise latent representations and their posterior estimates $p({\bf z}_x|{\bf{x}})$, $p({\bf z}_d|{\bf{d}})$, respectively. This is achieved by minimizing the following VAE loss function: 
\vspace{-0.2cm}
\begin{equation}
\small
 \begin{aligned}
 \setlength{\abovedisplayskip}{2pt}
   \mathcal{L}_{c} (\theta_x, \varphi_x; {\bf x}) &= {\mathbb E_{{\bf {x}} \sim p({\bf {x}})}} \{ D_{KL}\left({p({{\bf {z}}_x}|{\bf{x}})}||{q({\bf z}_x)}\right) \\
  & \quad - {\mathbb E_{{\bf {z}}_x \sim p({{\bf {z}}_x}|{\bf{x}})}} [\log {q({\bf x}|{\bf z}_x)} ]\},
   \end{aligned}
  \label{clean_vae}
  \setlength{\belowdisplayskip}{2pt}
\end{equation}
\begin{equation}
\small
 \begin{aligned}
 \setlength{\abovedisplayskip}{0pt}
   \mathcal{L}_{d} (\theta_d, \varphi_d; {\bf d}) &= {\mathbb E_{{\bf {d}} \sim p({\bf {d}})}} \{ D_{KL}\left({p({{\bf {z}}_d}|{\bf{d}})}||{q({\bf z}_d)}\right) \\
  & \quad - {\mathbb E_{{\bf {z}}_d \sim p({{\bf {z}}_d}|{\bf{d}})}} [\log {q({\bf d}|{\bf z}_d)} ]\},
   \end{aligned}
  \label{noise_vae}
\end{equation}
where $\theta_x, \varphi_x, \theta_d, \varphi_d$ {are} the DNN parameters for the related probability estimation \cite{xiang2022bayesian}. Additionally, NS-VAE is trained under the supervision of C-VAE and N-VAE's encoders. Based on the derivation in \cite{xiang2022bayesian}, the NS-VAE's training loss function is
\begin{equation}
\small
 \begin{aligned}
  & \mathcal{L}_{p} (\theta_y, \varphi_y; {\bf y}) \\
  & \quad = {\mathbb E_{{\bf {y}} \sim p(\bf {y}),{\bf {x}} \sim p(\bf {x})}} \{D_{KL}\left({p({\bf z}_x|{\bf{y}})}||{p({\bf z}_x|{\bf{x}})}\right) \\
  & \quad \quad + {\mathbb E_{{{\bf {z}}_x} \sim p({{\bf {z}}_x}|{\bf {y}})}}[\log \frac{p({\bf z}_x|{\bf x})}{q({\bf z}_x)}]\} \\
  & \quad \quad + {\mathbb E_{{\bf {y}} \sim p({\bf {y}}), {\bf {d}} \sim p({\bf {d}})}} \{D_{KL}\left({p({\bf z}_d|{\bf{y}})}||{p({\bf z}_d|{\bf{d}})}\right) \\
  &  \quad \quad + {\mathbb E_{{{\bf {z}}_d} \sim p({{\bf {z}}_d}|{\bf {y}})}}[\log \frac{p({\bf z}_d|{\bf d})}{q({\bf z}_d)}]\} \\
  & \quad \quad - {\mathbb E_{{\bf {y}} \sim p(\bf {y})}} \left[ {\mathbb E_{{{\bf {z}}_d,{\bf {z}}_x} \sim p({{\bf {z}}_d,{\bf {z}}_x}|\bf {y})}}\left[\log {q({\bf{y}}|{\bf z}_x,{\bf z}_d)} \right]\right],
   \end{aligned}
  \label{final_loss_funtion}
\end{equation}
where $\theta_y, \varphi_y$ are the NS-VAE's network parameters.

In the online SE stage, we assume that {the} ${{\bf z}_x, {\bf z}_d}$ sampled from $p({\bf z}_x|{\bf{x}})$ and $p({\bf z}_d|{\bf{d}})$ are approximately equal to the sample ${{\bf z}_x}$, ${{\bf z}_d}$ sampled from $p({\bf z}_x|{\bf{y}})$, $p({\bf z}_d|{\bf{y}})$, respectively. So, we separately use {the} NS-VAE encoder's two outputs as input of C-VAE and N-VAE to estimate related signals for SE.

\section{$\beta$-VAE-based speech enhancement}
\begin{figure}[!tbp]
  \centering
  \centerline{\includegraphics[scale=0.5]{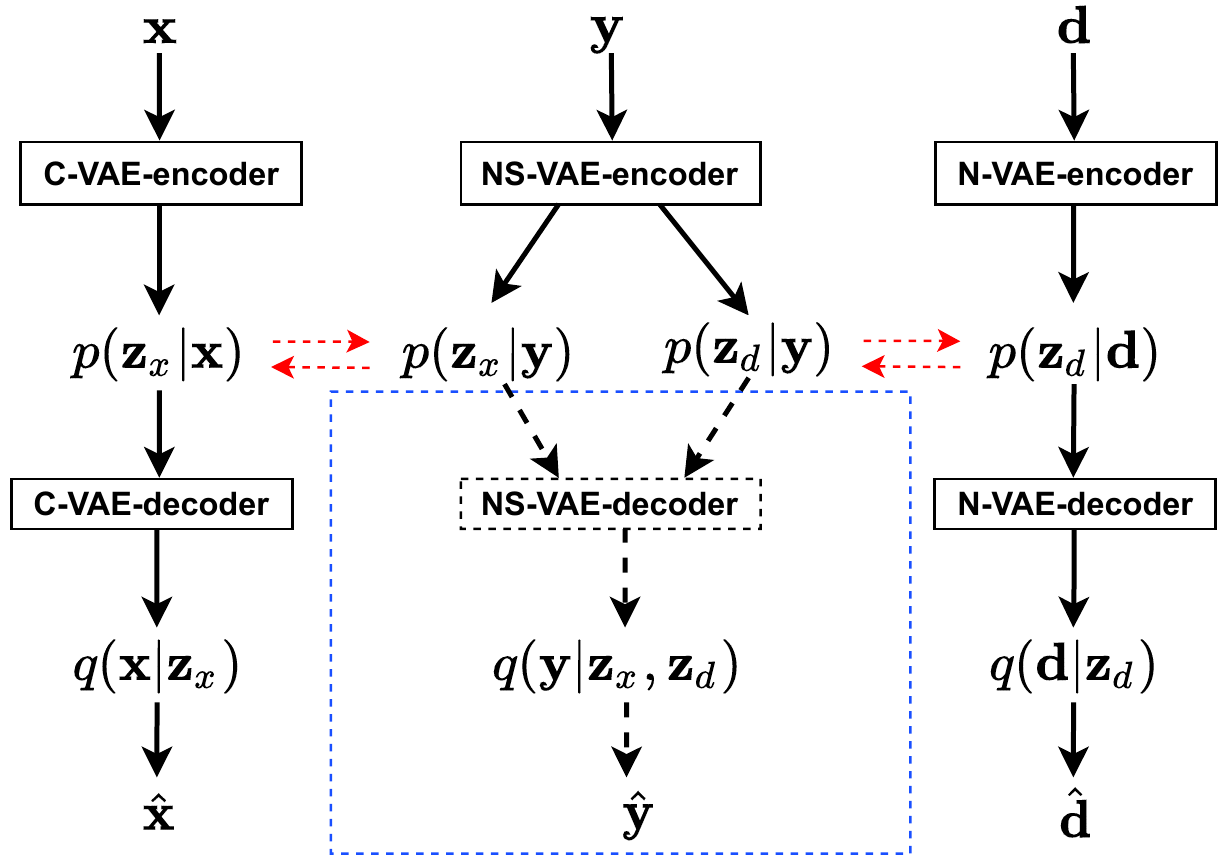}}
  \setlength{\abovecaptionskip}{0.0cm}
  \caption{Model illustration for PVAE and $\beta$-PVAE.}
  \label{fig:Bayesian_DNN}
  \vspace{-0.6cm}
\end{figure}
Inspired by $\beta$-VAE, we propose a novel $\beta$-VAE strategy (named $\beta$-PVAE) to further improve PVAE's SE performance. More specifically, $\beta$-VAE is used to improve PVAE's representation learning ability that can better disentangle speech and noise latent variables from the observed signal, which can help PVAE obtain better SE performance. In PVAE, all the PVAE's decoders are trained in an unsupervised way \cite{xiang2022bayesian}. The accuracy of the restored signal depends on the quality of latent representations. This means that the SE performance in PVAE is determined by the quality of speech and noise latent variables.

In \cite{xiang2022bayesian}, we derived a novel evidence lower bound (ELBO) (${\mathbb E_{{\bf {y}} \sim p(\bf {y})}[\log q(\bf {y})]} \ge -\mathcal{L}_{p}$). Additionally, $\beta$-VAE \cite{higgins2016beta} applies an adjustable hyperparameter $\beta$ in original VAE's \cite{kingma2013auto}  KL divergence term. Following $\beta$-VAE's property and PVAE's derivation \cite{xiang2022bayesian}, we apply this hyperparameter in the derived ELBO \cite{xiang2022bayesian}, the (\ref{final_loss_funtion}) can be written as  
\begin{equation}
\small
 \begin{aligned}
  & \mathcal{L}_{p} (\theta_y, \varphi_y; {\bf y}) \\
  & \quad = \beta{\mathbb E_{{\bf {y}} \sim p(\bf {y}),{\bf {x}} \sim p(\bf {x})}} \{D_{KL}\left({p({\bf z}_x|{\bf{y}})}||{p({\bf z}_x|{\bf{x}})}\right) \\
  & \quad \quad + {\mathbb E_{{{\bf {z}}_x} \sim p({{\bf {z}}_x}|{\bf {y}})}}[\log \frac{p({\bf z}_x|{\bf x})}{q({\bf z}_x)}]\} \\
  & \quad \quad + \beta{\mathbb E_{{\bf {y}} \sim p({\bf {y}}), {\bf {d}} \sim p({\bf {d}})}} \{D_{KL}\left({p({\bf z}_d|{\bf{y}})}||{p({\bf z}_d|{\bf{d}})}\right) \\
  &  \quad \quad + {\mathbb E_{{{\bf {z}}_d} \sim p({{\bf {z}}_d}|{\bf {y}})}}[\log \frac{p({\bf z}_d|{\bf d})}{q({\bf z}_d)}]\} \\
  & \quad \quad - \alpha{\mathbb E_{{\bf {y}} \sim p(\bf {y})}} \left[ {\mathbb E_{{{\bf {z}}_d,{\bf {z}}_x} \sim p({{\bf {z}}_d,{\bf {z}}_x}|\bf {y})}}\left[\log {q({\bf{y}}|{\bf z}_x,{\bf z}_d)} \right]\right].
   \end{aligned}
  \label{beta_final_loss_funtion}
\end{equation}
In (\ref{beta_final_loss_funtion}), we introduce a hyperparameter $\alpha$ in the restoration term. The purpose is to better analyze $\beta$-VAE \cite{higgins2016beta} in PVAE. Note, $\alpha$ will not generate any effects for the original $\beta$-VAE's property because what is important in (\ref{beta_final_loss_funtion}) is the weight ratio $\beta:\alpha$. This weight ratio can also be written as: $\gamma =\beta:\alpha = (\beta/\alpha):1$, which is equal to the original $\beta$-VAE's loss function in (\ref{beta_ELBO}). $\beta$-VAE \cite{higgins2016beta} indicates that a higher value of $\beta$ encourages VAE learning a more disentangled representation. Thus, we hypothesize that a higher value of $\beta:\alpha$ in (\ref{beta_final_loss_funtion}) can cause a better disentangling performance for speech and noise latent variables. This point will be verified by later experiments.

$\beta$-VAE usually has a trade-off problem between the disentanglement and signal reconstruction \cite{higgins2016beta}, which means that a good disentangled representation usually leads to poor signal reconstruction performance. In NS-VAE (as shown in Fig.~\ref{fig:Bayesian_DNN}), this trade-off is between the quality of observed signal reconstruction and the disentanglement of speech and noise latent variables. In SE application, we only need NS-VAE's disentanglement function, observed signal reconstruction is not useful (dashed part in Fig.~\ref{fig:Bayesian_DNN}). This means that we should set a very high weight ratio $\gamma$ to obtain a better disentanglement performance \cite{higgins2016beta}. Ideally, $\gamma \to +\infty$. One strategy to achieve this purpose is to set $\alpha = 0$, so the loss function (\ref{beta_final_loss_funtion}) can be rewritten as 
\begin{equation}
\small
 \begin{aligned}
 \setlength{\abovedisplayskip}{3pt}
  & \mathcal{L}_{\beta} (\theta_y; {\bf y}) = \beta{\mathbb E_{{\bf {y}} \sim p(\bf {y}),{\bf {x}} \sim p(\bf {x})}} \{D_{KL}\left({p({\bf z}_x|{\bf{y}})}||{p({\bf z}_x|{\bf{x}})}\right) \\
  & \quad \quad + {\mathbb E_{{{\bf {z}}_x} \sim p({{\bf {z}}_x}|{\bf {y}})}}[\log \frac{p({\bf z}_x|{\bf x})}{q({\bf z}_x)}]\} \\
  & \quad \quad + \beta{\mathbb E_{{\bf {y}} \sim p({\bf {y}}), {\bf {d}} \sim p({\bf {d}})}} \{D_{KL}\left({p({\bf z}_d|{\bf{y}})}||{p({\bf z}_d|{\bf{d}})}\right) \\
  &  \quad \quad + {\mathbb E_{{{\bf {z}}_d} \sim p({{\bf {z}}_d}|{\bf {y}})}}[\log \frac{p({\bf z}_d|{\bf d})}{q({\bf z}_d)}]\}.
   \end{aligned}
  \label{beta_pvae}
\end{equation}
In (\ref{beta_pvae}), it can be found that there is no reconstruction term. This means that we do not need to train the NS-VAE's decoder, which reduces the PVAE's training parameters. The dashed part in Fig.~\ref{fig:Bayesian_DNN} is removed in the proposed $\beta$-PVAE framework. Comparing the PVAE and proposed $\beta$-PVAE, we can find that the $\beta$-VAE can be used to optimize the PVAE's network structure and $\beta$-PVAE also addresses the $\beta$-VAE's trade-off problem for SE application. All in all, the combination of $\beta$-VAE and PVAE can not only improve PVAE's disentanglement performance, but also simplify its framework.

To summarize, the proposed $\beta$-PVAE includes a training and an enhancement stage for the SE application, which is similar to PVAE \cite{xiang2022bayesian}. In the training stage, C-VAE and N-VAE are separately pre-trained by self-supervision using (\ref{clean_vae}) and (\ref{noise_vae}). After that, we apply (\ref{beta_pvae}) to train NS-VAE. In the enhancement stage, we can separately use {the} NS-VAE encoder's two outputs as input of C-VAE and N-VAE to obtain the prior distributions $q({\bf{x}}|{\bf z}_x)$ and $q({\bf{d}}|{\bf z}_d)$ for SE. Moreover, to calculate (\ref{beta_pvae}), related prior and posterior distributions need to be determined. Here, all the estimations of these distributions are the same as PVAE. More details can be found in \cite{xiang2022bayesian}.

\section{Experiments}

In this section, we report two experiments. First, we will investigate the disentanglement ability of the latent variables in the proposed algorithm. In addition, $\beta$-PVAE's SE performance will be indicated. 

{\bf Datasets:} In this work, we use the DNS challenge 2021 corpus \cite{reddy2021interspeech} to evaluate the performance of the proposed algorithm. We select English speakers
and randomly split 70\% of speakers for training, 20\% for validation, and 10\% for evaluation. Then, all the noise from the DNS noise corpus are randomly divided into training, validation, and test noise in a proportion similar to that used for speech utterances. Next, the corresponding  training, validation, and test corpus for speech and noise are randomly mixed using DNS script \cite{reddy2021interspeech} with random signal-to-noise ratio (SNR) levels (SNR range is from --10dB to 15dB). Other parameters of signal mixing are the default values in the DNS script \cite{reddy2021interspeech}. Finally, we randomly choose 20 hours mixed training utterances, 5 hours mixed validation utterances, and 1 hour mixed test utterances to build experimental dataset. All signals are down-sampled to 16 kHz.  

\begin{table}[!t]
 \centering
 \setlength{\abovecaptionskip}{0.08cm}
  \caption{Average STOI, PESQ, and SI-SDR comparison for $\beta$-PVAE under different $\gamma$ with a 95\% confidence interval ($\beta$-PVAE is equal to PVAE when $\gamma=1$) }
  \label{tab: all algorithm}
  \centering
    \begin{tabular}{cccc}
    \toprule
    Method & STOI & PESQ &SI-SDR \\
    \midrule
    Noisy &88.94($\pm\,$1.77)&2.29($\pm\,$0.02) &8.36($\pm\,$1.13) \\
    \cmidrule(lr){1-4}
    Oracle &98.12($\pm\,$0.35) &4.19($\pm\,$0.00) &19.84($\pm\,$0.92) \\
    \cmidrule(lr){1-4}
    PVAE ($\gamma=1$) &89.33($\pm\,$1.72) &2.59($\pm\,$0.03) &10.31($\pm\,$1.03) \\
    \cmidrule(lr){1-4}
     $\gamma=2$ &89.81($\pm\,$1.67) &2.69($\pm\,$0.02) &11.84($\pm\,$0.97) \\
    \cmidrule(lr){1-4}
    $\gamma=5$ &89.76($\pm\,$1.64) &2.70($\pm\,$0.02) &12.23($\pm\,$0.93) \\
    \cmidrule(lr){1-4}
    $\gamma=10$  &89.94($\pm\,$1.70) &2.71($\pm\,$0.02) &12.31($\pm\,$0.94) \\
    \cmidrule(lr){1-4}
    $\gamma=100$  &89.98($\pm\,$1.70) &2.72($\pm\,$0.02) &12.45($\pm\,$0.94) \\
     \cmidrule(lr){1-4}
    $\gamma=1000$  &90.02($\pm\,$1.71) &2.74($\pm\,$0.01) &12.55($\pm\,$0.94) \\
     \cmidrule(lr){1-4}
    $\gamma= +\infty$  &90.05($\pm\,$1.71) &2.75($\pm\,$0.01) &13.20($\pm\,$0.95) \\
    \bottomrule                             
  \end{tabular}
  \vspace{-0.6cm}
\end{table}

{\bf Experimental settings:} In the experiments, the neural structures for C-VAE and N-VAE are the same. Their encoders include four hidden 1D convolutional layers \cite{ luo2019conv}. The number of channels in each layer is 32, 64, 128, and 256. The size of each convolving kernel is 3. The two output layers of the encoders are fully connected layers with 128 nodes. Their decoders consist of four hidden 1D convolutional  layers (the channel number of each layer is 256, 128, 64, and 32 with 3 kernel) and two fully connected output layers with 257 nodes. For NS-VAE, its encoder's hidden layer setting is the same as C-VAE. NS-VAE's encoder has four output layers with 128 nodes. For C-VAE, N-VAE, and NS-VAE, their activation functions in the hidden and output layer are ReLU and linear activation function, respectively. All networks are trained by the Adam algorithm with a 128 mini-batch size. 


\begin{figure}[!tbp]
  \centering
  \centerline{\includegraphics[scale=0.5]{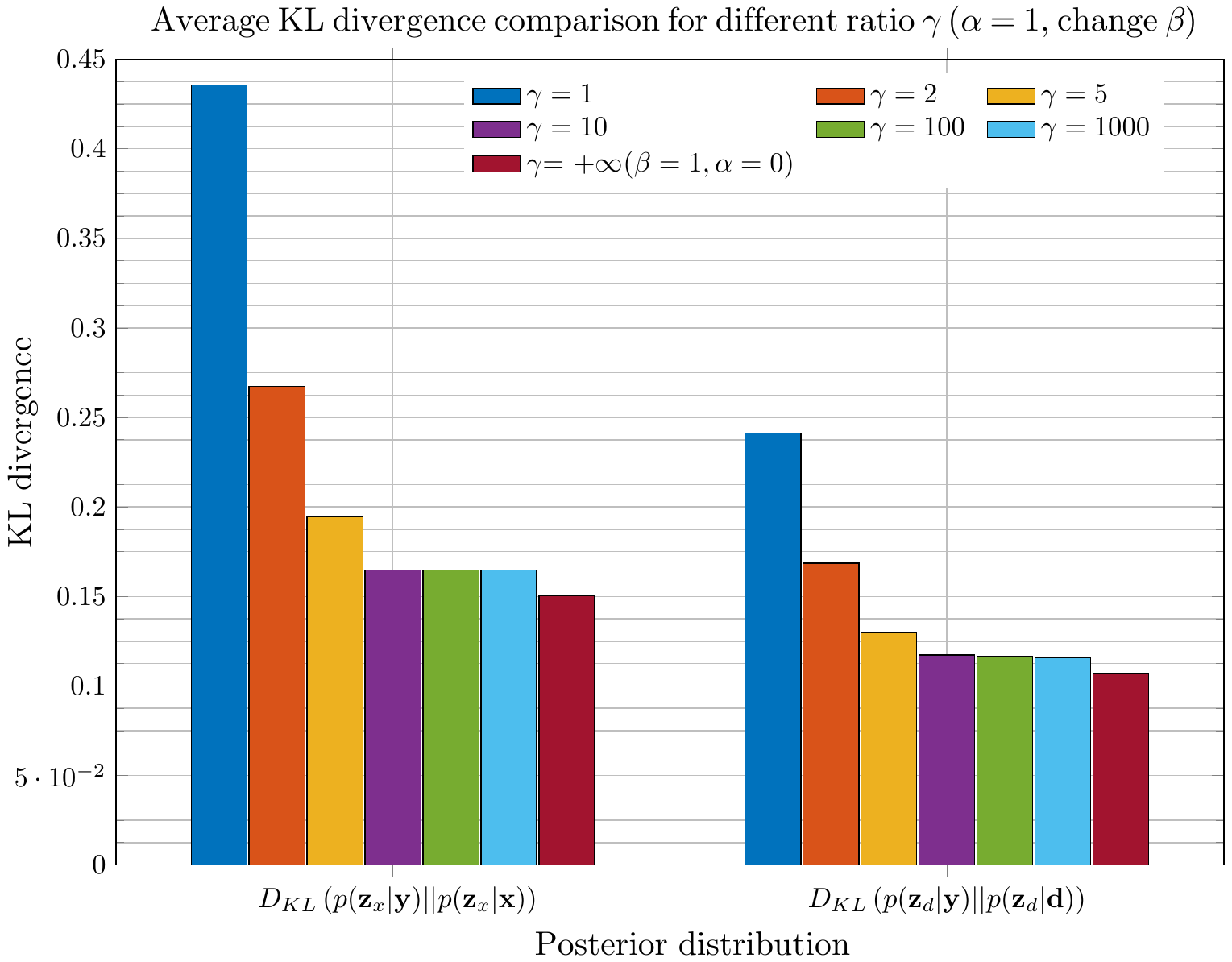}}
  \setlength{\abovecaptionskip}{0.0cm}
  \caption{Average KL divergence comparison of the posterior distribution for different ratio $\gamma$.}
  \label{fig:KL}
  \vspace{-0.8cm}
\end{figure}

{\bf Experimental results:} To evaluate the SE performance of various algorithms, we will use scale-invariant signal-to-distortion ratio (SI-SDR) in decibel (dB) \cite{le2019sdr}, short-time objective intelligibility (STOI)\cite{taal2011algorithm}, and perceptual evaluation of speech quality (PESQ)\cite{rix2001perceptual} as evaluation metrics. 

First, we will investigate $\beta$-PVAE's performance in disentangling speech and noise latent variables. Based on our previous derivation and analysis \cite{xiang2022bayesian}, $\beta$-PVAE's SE performance is determined by disentanglement performance. Table.~\ref{tab: all algorithm} 'Oracle' shows the SE performance with a 95\% confidence interval if latent variables are completely disentangled. Here, the signals are reconstructed by mask estimation \cite{wang2014training}. The complete disentanglement means that they 
have the same posterior forms: $p({\bf z}_x|{\bf{x}}) = p({\bf z}_x|{\bf{y}})$ and $p({\bf z}_d|{\bf{d}}) = p({\bf z}_d|{\bf{y}})$. This is because $p({\bf z}_x|{\bf{x}})$ and $p({\bf z}_d|{\bf{d}})$ are learned in an unsupervised way with speech or noise only, which ensures that their latent representation only contains speech or noise representation. 'Oracle' results indicate that $\beta$-PVAE achieves a very satisfactory SE performance in SI-SDR, STOI, and PESQ, which shows the importance of disentangling latent variable for achieving excellent SE performance. The NS-VAE's purpose is to disentangle different representations from the observed signal and obtain the closest possible speech and noise posterior. Next, we use KL divergence to evaluate the practical disentanglement performance in latent space. A better disentanglement can lead to a lower KL divergence ($D_{KL}\left({p({\bf z}_d|{\bf{y}})}||{p({\bf z}_d|{\bf{d}})}\right)$ and $D_{KL}\left({p({\bf z}_x|{\bf{y}})}||{p({\bf z}_x|{\bf{x}})}\right)$). Fig.~\ref{fig:KL} shows the average KL divergence comparison of validation samples for using different ratios $\gamma = \beta:\alpha$ in loss function (\ref{beta_final_loss_funtion}) to train NS-VAE. In (\ref{beta_final_loss_funtion}), we keep $\alpha = 1$ and change different $\beta$ to determine ratio $\gamma$, and $\gamma = +\infty$ means that $\alpha = 0, \beta = 1$, which is equal to (\ref{beta_pvae}). In Fig.~\ref{fig:KL}, we see that the KL divergence decreases with the increase of $\gamma$ for both speech and noise latent variables, which means that the disentangled posteriors get closer to the true posteriors and the NS-VAE achieves a better disentanglement performance. When NS-VAE's decoder is removed ($\gamma = +\infty, \alpha = 0, \beta = 1$), NS-VAE can acquire the best posterior estimation. This verifies our hypothesis and deduction in Section 3. Additionally, although we have an inaccurate posterior conditional assumption $p({\mathbf{z}_x},{\mathbf{z}_d}|\mathbf{y}) = p(\mathbf{z}_x|\mathbf{y})p(\mathbf{z}_d|\mathbf{y})$,  Fig.~\ref{fig:KL} shows that NS-VAE can still estimate a satisfactory posterior with a low KL divergence. However, this inaccurate assumption may hinder NS-VAE 
from obtaining a lower KL divergence when $\gamma = +\infty$.

Next, we will evaluate the SE performance of the proposed $\beta$-PVAE. We use basic PVAE \cite{xiang2022bayesian} as the reference method, which can be more direct to find the effects of $\beta$-VAE for the previous PVAE. The enhanced speech is obtained by mask estimation \cite{wang2014training}. Table.~\ref{tab: all algorithm} shows the experimental results. We find that $\beta$-PVAE achieves a very significant STOI, PESQ, and SI-SDR improvement over PVAE (from $\beta=1$ to $\beta=2$ ). This indicates that good disentanglement performance in latent space can directly lead to an improvement in speech quality and intelligibility. In addition, $\beta$-PVAE achieves the best SE performance when $\beta=+\infty$. This illustrates that the proposed $\beta$-PVAE can effectively improve PVAE's SE performance with a simpler network structure.
\vspace{-0.1cm}
\section{Conclusions}
\vspace{-0.1cm}

In this paper, a $\beta$-PVAE-based SE method is proposed to improve previous PVAE's SE performance. More specifically, $\beta$-PVAE can improve PVAE's ability to disentangle speech and noise latent variables from the observed signal. In addition, based on VAE's application in SE, the proposed $\beta$-PVAE addresses the trade-off problem between disentanglement and signal reconstruction, which widely exists in $\beta$-VAE. Compared with the previous PVAE algorithm, $\beta$-PVAE also simplifies its neural network and reduces the number of training parameters when improving the SE performance. Experimental results indicate that a good signal representation can achieve a very satisfactory SE performance. Moreover, $\beta$-PVAE obtains a better disentanglement performance and achieves higher SI-SDR, PESQ, and STOI scores than PVAE. In future work, we believe that $\beta$-PVAE can achieve better SE performance by improving the latent space disentanglement performance or the decoder's signal reconstruction ability.

\bibliographystyle{IEEEtran}
\bibliography{IEEEabrv,myabrv_new,my_reference}

\vspace{12pt}

\end{document}